\documentstyle[seceq,epsf,wrapft,supplement]{ptptex}

\newcommand{\calHz}{{\cal{H}}_0}
\newcommand{\vecr}{\mbox{\boldmath $r$}}
\newcommand{\vecrs}{\mbox{\boldmath $r$}\sigma}
\newcommand{\vecrst}{\mbox{\boldmath $r$}\tilde{\sigma}}
\newcommand{\vecrsp}{\mbox{\boldmath $r$}'\sigma'}

\newcommand{\vecrspt}{\mbox{\boldmath $r$}'\tilde{\sigma}'}

\newcommand{\vphi}{\varphi}

\newcommand{\htl}{\tilde{h}}

\newcommand{\phibari}{\overline{\phi}_{\tilde{i}}}

\newcommand{\rhot}{\tilde{\rho}}
\newcommand{\psid}{\psi^\dag}

\newcommand{\calA}{{\cal{A}}}
\newcommand{\calB}{{\cal{B}}}
\newcommand{\calGz}{{\cal{G}}_{0}}

\newcommand{\eps}{\epsilon}
\newcommand{\Tr}{{\rm Tr}}

\title{
Collective Excitations and Pairing Effects
in Drip-Line Nuclei
}
\subtitle{Continuum RPA in Coordinate-Space HFB}

\author{
Masayuki {\sc Matsuo}
\footnote{E-mail address:
matsuo@nt.sc.niigata-u.ac.jp}
}

\inst{
Graduate School of Science and Technology,
Niigata University, Niigata, 950-2181, Japan}

\recdate{
}

\abst{
We discuss novel features of a new continuum RPA
formulated in the coordinate-space Hartree-Fock-Bogoliubov
framework. This continuum quasiparticle RPA 
takes into account both the one- and two-particle escaping channels.
The theory is tested with 
numerical calculations for monopole, dipole and quadrupole 
excitations in neutron-rich oxygen isotopes near the drip-line.
Effects of the particle-particle RPA correlation 
caused by the pairing interaction
are discussed in detail, and importance of
the selfconsistent treatment is emphasized.

}

\begin{document}

\maketitle

\section{Introduction}
Collective excitation in unstable nuclei is one of the most attractive
subjects since the exotic structures in the ground state, 
such as  halo, skin, and the presence of loosely bound nucleons,
may cause new features in the excitations, e.g.
the low-energy dipole mode that is being discussed extensively.
The random phase approximation (RPA) or the linear response theory
is one of the most powerful framework to investigate such problems
microscopically. 
Indeed the continuum RPA theory 
in the coordinate-space representation\cite{Shlomo,Bertsch} has played
major roles so far since it can describe the continuum
states crucial for nuclei near drip-line.\cite{HaSaZh,FaCRPA,Colo-1}

The pairing correlation is another key feature of 
drip-line nuclei.\cite{DobHFB,Hansen,Esbensen} 
To treat the coupling of the continuum
states as well as the density dependence of the pairing
correlation, 
the Hartree-Fock-Bogoliubov (HFB)
theory formulated in the coordinate-space representation
\cite{DobHFB,Belyaev} has been developed 
while the conventional BCS approximation has inherent deficiency.

It is therefore important to combine the continuum RPA and 
the coordinate space HFB in a consistent way 
in order to describe the excitations in unstable nuclei 
near drip-line,
especially when the pairing correlation play crucial roles.
We have recently shown that a new quasiparticle RPA (QRPA) satisfying
this requirement is indeed possible.\cite{Matsuo}
In the present paper, we discuss 
characteristic features of the theory and analyze excitations of
near-drip-line nuclei, focusing on effects of the pairing,
by using numerical calculations performed 
for the monopole, dipole and quadrupole excitations 
in neutron-rich oxygen isotopes. 
The previous continuum QRPA approaches
employ the BCS approximation.\cite{Pl88FFSth,Bo96FFSGT,Kam98FFSe1,Hagino}
Other QRPA approaches applied to unstable nuclei 
neglect the escaping effects since some use the
BCS quasiparticle basis,\cite{Khan-Giai,Khan-etal,Colo-2}
and other adopt the coordinate-space HFB but
use the discretized canonical basis.\cite{Engel}
The present formalism provides
the first consistent continuum QRPA approach for the drip-line nuclei
with pairing correlation.

\section{Continuum RPA in coordinate-space HFB}

The Hartree-Fock-Bogoliubov theory describes
the pairing correlation 
in terms of quasiparticles and selfconsistent mean-fields including 
the pairing potential. 
To correctly describe behavior of the quasiparticles
in the surface and
exterior regions related to halo or skin, it is preferable to
solve the HFB equation in the coordinate-space 
representation\cite{DobHFB,Belyaev}
\begin{equation}\label{grHFB}
\calHz \phi_i(\vecrs) = E_i \phi_i(\vecrs),
\end{equation}
which determines the quasiparticle states 
and the associated two-component wave functions
\begin{equation} \label{spinor}
\phi_i(\vecrs) \equiv 
\left(
\begin{array}{c}
\vphi_{1,i}(\vecrs) \\
\vphi_{2,i}(\vecrs)
\end{array}
\right).
\end{equation}
Here the HFB mean-field Hamiltonian is expressed in a $2 \times 2$
matrix form
\begin{equation} \label{Hamlmat}
\calHz(\vecrs,\vecrsp) \equiv 
\left(
\begin{array}{cc}
h(\vecrs,\vecrsp) -\lambda\delta(\vecr-\vecr')\delta_{\sigma\sigma'} & 
                                    \htl(\vecrs,\vecrsp) \\
\htl^*(\vecrst,\vecrspt) & 
  -h^*(\vecrst,\vecrspt)+\lambda\delta(\vecr-\vecr')\delta_{\sigma\sigma'}
\end{array}
\right)
\end{equation}
where $h$ includes the kinetic energy and the Hartree-Fock 
field in the particle-hole (ph) channel and $\htl$ 
is the selfconsistent pairing field in the particle-particle (pp) channel. 
They are expressed in terms of the effective two-body interactions and the
normal and pair densities although 
we omit here their detailed expression. Properties of
the static HFB equations and techniques to solve them 
are known.\cite{DobHFB,Belyaev}
The quasiparticle excitation energy $E_i$ is defined with respect to
the Fermi energy $\lambda (<0)$. The spectrum
becomes continuous for $E_i > |\lambda|$ and the quasiparticles
above the threshold energy $|\lambda|$ can escape from the nucleus. 
This is a special feature we have to take care of when we describe
weakly bound systems with pairing correlation.

In order to describe the linear response of the system, we need to know 
the motion of two quasiparticles propagating under the
HFB mean-field Hamiltonian. 
Assuming that the external
field and the selfconsistent field are the local one-body
fields expressed in terms of the normal density 
$\rho(\vecr)=\sum_\sigma \psid(\vecrs)\psi(\vecrs) $ 
and the pair densities $\rhot_\pm(\vecr)=
{1\over 2}\sum_\sigma
\left(\psi(\vecrst)\psi(\vecrs)\pm\psid(\vecrs)\psid(\vecrst)\right)$, 
it is enough to consider response function for these operators. 
The unperturbed response function $R_0(\omega)$ at frequency $\omega$, 
that neglects effect of the residual
interaction, is easily derived from a time-dependent extension of 
Eq.(\ref{grHFB}) as 
\begin{eqnarray}\label{uresp1}
R_{0}^{\alpha\beta}(\vecr,\vecr',\omega) ={1\over2}\sum_i \sum_{\sigma\sigma'}
&&\left\{
 \phibari^{\dag}(\vecrs)\calA\calGz(\vecrs,\vecrsp,-E_i+\hbar\omega+i\eps)
\calB\phibari(\vecrsp) \right. \nonumber\\
&& \hspace{-10mm} \left. + 
\phibari^{\dag}(\vecrsp)\calB\calGz(\vecrsp,\vecrs,-E_i-\hbar\omega-i\eps)
\calA\phibari(\vecrs)\right\} 
\end{eqnarray}
with use of 
the HFB Green function $\calGz(E+i\eps)=\left(E+i\eps-\calHz\right)^{-1}$
and the wave functions $\phibari(\vecrs)$ of
the quasiparticle states. Here $\phibari(\vecrs)$ is the one associated
with the negative energy quasiparticle state (with energy
$-E_i$) conjugate to a positive energy state $\phi_i(\vecrs)$ with $E_i$. 
The index $\alpha,\beta$ and
the symbol $\calA, \calB$ refer to the three
kinds of densities $\rho,\rhot_+$ and $\rhot_-$.
Structure of Eq.(\ref{uresp1}) is similar 
to the familiar form for unpaired systems\cite{Shlomo} 
except that we here use the quasiparticle
wave functions and the HFB Green function. 
However this expression is not
satisfactory since the summation $\sum_i$ 
assumes that all the quasiparticle states
belong to a discrete spectrum. 
We have to take a continuum limit of this summation
to deal with the continuum effects accurately.

The continuum limit is introduced by first replacing 
the summation with a contour integral of the
HFB Green function $\calGz(E)$ in the complex $E$ plane. The contour
is chosen so that it encloses all the quasiparticle poles at $-E_i$
of the negative energy quasiparticle states (see, Fig.1). By doing this, the
unperturbed response function can be expressed as
\begin{eqnarray}\label{uresp3}
R_{0}^{\alpha\beta}(\vecr,\vecr',\omega) =
{1\over 4\pi i}\int_C dE \sum_{\sigma\sigma'}
&&\left\{ 
\Tr\calA\calGz(\vecrs,\vecrsp,E+\hbar\omega+i\eps)
        \calB\calGz(\vecrsp,\vecrs,E)  \right. \nonumber\\
&&\hspace{-10mm} \left. +\Tr\calA\calGz(\vecrs,\vecrsp,E)
        \calB\calGz(\vecrsp,\vecrs,E-\hbar\omega-i\eps) \right\}
\end{eqnarray}
in terms of products of the HFB Green function $\calGz$ 
and their integral. Now 
we can implement the continuum quasiparticle spectrum by
adopting the exact HFB Green function which satisfies 
the proper boundary condition 
of outgoing wave for the continuum states.\cite{Belyaev} 
The exact HFB Green function
has poles on the real $E$ axis in the interval $ -|\lambda| < E <
|\lambda|$, corresponding to the discrete bound quasiparticle states,
and has the branch cuts, associated with the continuum states, 
along the real axis for $E>|\lambda|$ and $E<-|\lambda|$ (Fig.1). 
We incorporate in this way  the continuum states within the framework.

\begin{figure}[h]
\centerline{\epsfxsize = 80mm\epsfbox{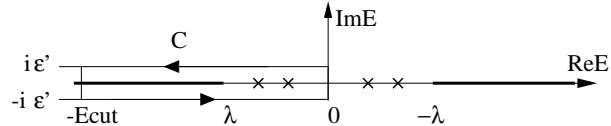}}
\caption{The contour $C$ in the integral representation of the
response function. The crosses represent the poles at
$E=\pm E_i$ corresponding to the discrete bound quasiparticle states. The
thick lines are the branch cuts associated with the continuum states.
The imaginary part $\eps'$ must satisfy the condition $0<\eps' < \eps$.}
\label{fig:1}
\end{figure}

Having the continuum unperturbed response function $R_0(\omega)$, 
we are then able to take into account the RPA correlation 
caused by the residual interactions. The RPA linear response equation
for the transition densities at frequency $\omega$  reads
\begin{equation} \label{rpa}
\left(
\begin{array}{c}
\delta\rho(\vecr,\omega) \\
\delta\rhot_+(\vecr,\omega) \\
\delta\rhot_-(\vecr,\omega) 
\end{array}
\right)
=\int d\vecr'
\left(
\begin{array}{ccc}
& & \\
& R_0^{\alpha\beta}(\vecr,\vecr',\omega)& \\
& & 
\end{array}
\right)
\left(
\begin{array}{l}
\kappa_{ph}(\vecr')\delta\rho(\vecr',\omega) + v^{ext}(\vecr') \\
\kappa_{pp}(\vecr')\delta\rhot_+(\vecr',\omega) \\
-\kappa_{pp}(\vecr')\delta\rhot_-(\vecr',\omega)
\end{array}
\right)
\end{equation}
where $\kappa_{ph}(\vecr)$ and $\kappa_{pp}(\vecr)$ are 
the residual interactions in the ph- and pp-channels. 
The strength function for the external one-body field
$v^{ext}(\vecr)$ is evaluated as  
$f(\omega)=
- {1 \over \pi}{\rm Im}\int d\vecr v^{ext}(\vecr)^*\delta\rho(\vecr,\omega)$. 

An important feature of the present continuum QRPA is that 
the linear response equation (\ref{rpa}) takes into account 
the pair transition density
$\delta\rhot_\pm$ as well as 
the normal transition density $\delta\rho$. Note that the three 
transition densities couple since the system has pairing correlation.
By solving Eq.(\ref{rpa}), the RPA correlations 
responsible for the excited states, represented by the ring
diagram, are taken into account both 
in the ph-channel (through $\delta\rho$) and in
the pp-channel ($\delta\rhot_\pm$).
The particle-particle RPA correlation may be called the
{\it dynamical pairing correlation} to distinguish from the
pairing correlation already taken into account 
as the quasiparticles in the HFB description of the 
ground state. We demonstrate in the following important roles played
by the dynamical pairing correlation for excitations of
nuclei near drip-line.

Another novel aspect of the theory concerns with 
the continuum states. The two-quasiparticle states appearing in the
QRPA formalism are classified in three groups;
the first consists of two nucleons both occupying the bound
discrete quasiparticle states, the second with one particle
 in the bound states and the other in the continuum states, 
and the last with two particles both in the continuum states. 
The three configurations are incorporated through 
the product of two HFB Green function $\calGz$ appearing
in Eq.(\ref{rpa}) 
since $\calGz$ describes exactly both the discrete and continuum
states. Thus 
the present linear response theory 
contains both the channel of one-nucleon escaping 
and that of two-nucleon emission. The threshold energy for
the two-nucleon channel is twice the Fermi energy $E_{th,2}=2|\lambda|$,
while the threshold for one-nucleon escape is $E_{th,1}=|\lambda|+E_{i,min}$
where $E_{i,min}$ is the lowest of the quasiparticle energy $E_i$.
See Ref.\citen{Matsuo} for details of the formalism.

\section{Numerical analysis for oxygen isotopes}

\subsection{Monopole: pairing selfconsistency}

In the following, we present our numerical analysis performed for 
the neutron-rich even-even oxygen isotopes including the neutron
drip-line nucleus
$^{24}$O. The adopted model assumes the Woods-Saxon
potential for the single-particle potential. As for the residual interaction
in the ph-channel, the Skyrme-type density-dependent delta
force 
$
v_{ph}(\vecr,\vecr')=
\left(t_0(1+x_0P_\sigma)+t_3(1+x_3P_\sigma)\rho(r)\right)
\delta(\vecr-\vecr')
$
is used. The model parameters for the Woods-Saxon
potential and the Skyrme force are taken the same 
as Shlomo-Bertsch.\cite{Shlomo} (Note that the adopted Woods-Saxon
parameters are slightly different from those in Ref.\citen{Matsuo}.) 
Although this modeling of the potential and the ph-interaction 
is not selfconsistent, an approximate selfconsistency
is satisfied 
by renormalizing the interaction strength $t_0$ and $t_1$
with an overall factor $f$ so that the dipole response has a zero-energy
mode corresponding to the spurious center of mass motion.\cite{Shlomo} 

For the pairing interaction, we adopt the density-dependent
delta force\cite{Esbensen,Te95HFBsd}
\begin{equation}\label{ddpair}
v_{pair}(\vecr,\vecr')={1\over2}V_0(1-P_\sigma)
(1-\rho(r)/\rho_0)\delta(\vecr-\vecr'),
\end{equation}
with $\rho_0=0.16\ {\rm fm}^{-3}$.
We use the same force to obtain the HFB pairing
field for the ground state 
and also to solve the linear response equation for the excitations.
Thus the calculation is selfconsistent in the pp-channel.

The pairing force strength $V_0$ is fixed to a value 
$V_0=520\ {\rm fm}^{-3}$MeV
which gives an average neutron pairing gap $\left<\Delta\right>$ 
(see Ref.\citen{Matsuo}for its definition) reproducing the global trend 
$\left<\Delta\right>\approx {12/\sqrt{A}}$ MeV in $^{18,20}$O. 
The calculated value is 
$\left<\Delta\right>=2.37, 2.83, 2.84, 2.89$ MeV for neutron and
zero gap for proton in $^{18-24}$O.  The model parameters and
the procedure of calculation is the same as those in Ref.\citen{Matsuo}
except the small difference in the Woods-Saxon parameters and the
force renormalization factor, which is $f=0.689, 0.704, 0.750, 0.775$
for $^{18-24}$O. Using the
present parameters, the peak energies of the 
giant quadrupole and dipole resonances are lifted
by about a few MeV (cf. Fig.3 compared with Fig.2 in
Ref.\citen{Matsuo}),  giving better description of GR's
in $^{16}$O. The calculation will be
further improved if we use the Hartree-Fock potential
for the ph-part. We will not discuss here quantitative (dis)agreement
with the experiments,\cite{Khan-etal,Jewell,Thirolf,Azaiez,GSI} 
but rather focus on qualitative aspects
seen in the theoretical analysis.
The small imaginary part in the response function is fixed to $\eps=0.2$ MeV.
It has an effect to
bring an additional width of 0.4 MeV in 
the calculated strength function. 

\begin{wrapfigure}{r}{6.6cm}
\centerline{\epsfxsize = 66mm\epsfbox{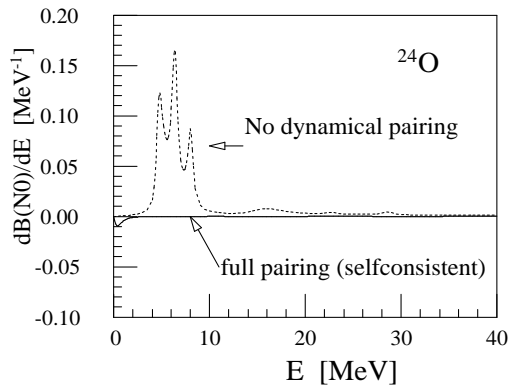}
}
\caption{The strength function for the neutron number operator in
$^{24}$O.} 
\label{fig:2}
\end{wrapfigure}

Let us first emphasize 
importance of the selfconsistent treatment of the pairing 
in the linear response equation by looking into the
Nambu-Goldstone mode associated with the nucleon number conservation.
Fig.2 shows the monopole strength function for the neutron number 
operator $\hat{N}=\int d\vecr\sum_\sigma \psid(\vecrs)\psi(\vecrs)$. 
There should be no response to the number operator
since it is the conserved quantity. The calculated strength function
(solid line) exhibits this feature, and 
there is essentially no spurious excitation that could have
been caused by the nucleon number
mixing in the HFB ground state. 
Note that the spurious strength for $\hat{N}$ were induced 
if we took into account the pairing correlation only 
in the static mean-field and neglected the dynamical pairing correlation 
i.e. neglecting the pairing interaction in the
linear response equation (\ref{rpa}) (see the
dotted curve in Fig.2). In the selfconsistent calculation
the Nambu-Goldstone mode has an excitation energy
very close to zero. We can shift this excitation energy to exact
zero by modifying the pairing force strength $V_0$ in Eq.(\ref{rpa})
just by less than 1\%, indicating that a good accuracy 
for the pairing selfconsistency is obtained in the actual calculation.

\subsection{Quadrupole excitation}

\begin{figure}
\centerline{\epsfxsize = 80mm\epsfbox{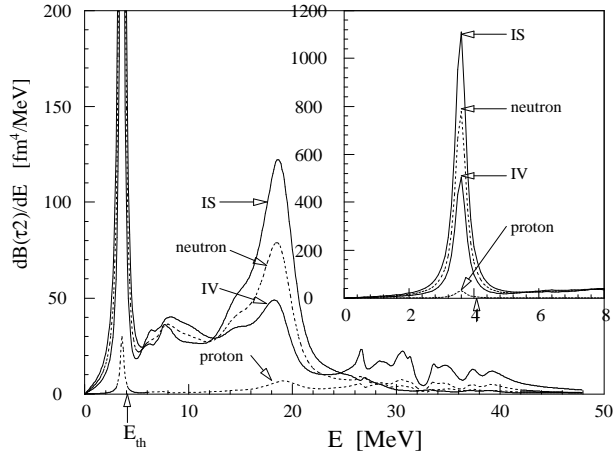}
}
\caption{The strength functions for the isoscalar, isovector,
proton and neutron quadrupole moments 
in $^{24}$O. The threshold energy $E_{th,1n}=E_{th,2n}=2|\lambda|=4.08$ MeV 
for neutron emission is indicated by an arrow.
}
\label{fig:3}
\end{figure}

The strength functions for the 
proton, neutron, isoscalar, and isovector quadrupole operators
calculated for the drip-line nucleus $^{24}$O are shown in Fig.3.
The peak 
around $E\approx 18$ MeV in the isoscalar strength and the broad
distribution 
\begin{wrapfigure}[27]{r}{6.6cm}
\centerline{
 \epsfxsize = 50mm\epsfbox{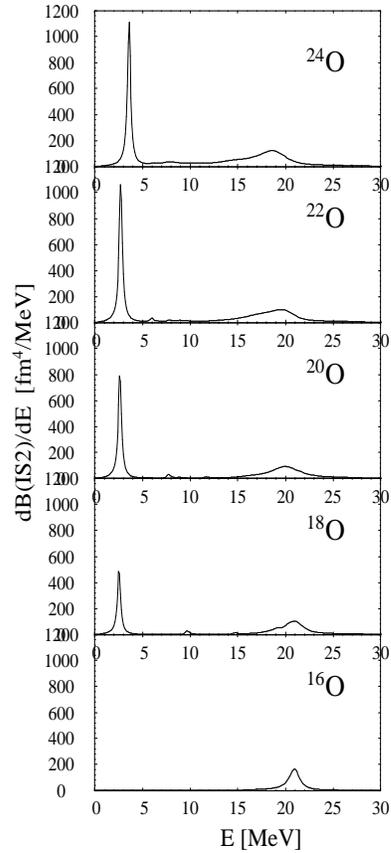}
}
\caption{Isoscalar quadrupole strength function for $^{16-24}$O.}
\label{fig:4}
\end{wrapfigure}
around $E\approx 25-40$ MeV of the isovector 
strength 
correspond to the isoscalar and isovector giant quadrupole 
resonances, respectively,
although isoscalar and isovector characters are mutually mixed.   
The most prominent feature is presence of the intense low-lying state
at $E=3.6$ MeV, which is very close to the threshold energy 
$E_{th,1n}=E_{th,2n}=4.08$ MeV
for the one- and two-neutron escape. 
This low-lying $2^+$ state has 
large neutron strength which is overwhelming the proton strength. The energy
weighted sum of the neutron (and the isoscalar) strength in this state
amounts to 12 (13)\% of the sum-rule value. 
The neutron transition density $\delta\rho(r)$ has a large 
peak in the surface region. Thus this low-lying state 
has a characteristic of neutron surface vibration;
for which the dominance of neutron
strength marks significant difference from
the isoscalar low-lying $2^+$ states in the stable nuclei. 
(The slightly lower excitation energy and the weaker strength
of the low-lying $2^+$ state in comparison with the calculation
in Ref.\citen{Matsuo} are due to the different Woods-Saxon parameter.)

Figure 4 shows 
systematics of the isoscalar quadrupole strength for $^{16-24}$O.
The isoscalar strength (and the neutron strength not shown here) of
the low-lying $2^+$ state increases with the neutron number.
The $B({\rm E}2)$ value of this state, on the other hand, 
is not very dependent of the neutron number;
$B({\rm E}2)=17, 18, 20, 18 \ e^2{\rm fm}^4$ for $A=18-24$. 
Thus the neutron character of this vibration mode
is enhanced as  the drip-line is approached.
Significant amount of neutron strength in the interval 
between the low-lying $2^+$ state and the GQR, seen in 
the near-drip-line nuclei $^{22,24}$O, arises from the
neutron continuum states.
It is also seen that the width of the isoscalar 
giant quadrupole resonance increases with the neutron number.
This can be naturally interpreted as an increase
of the neutron escaping width.

\begin{figure}[t]
\centerline{
\epsfxsize = 120mm\epsfbox{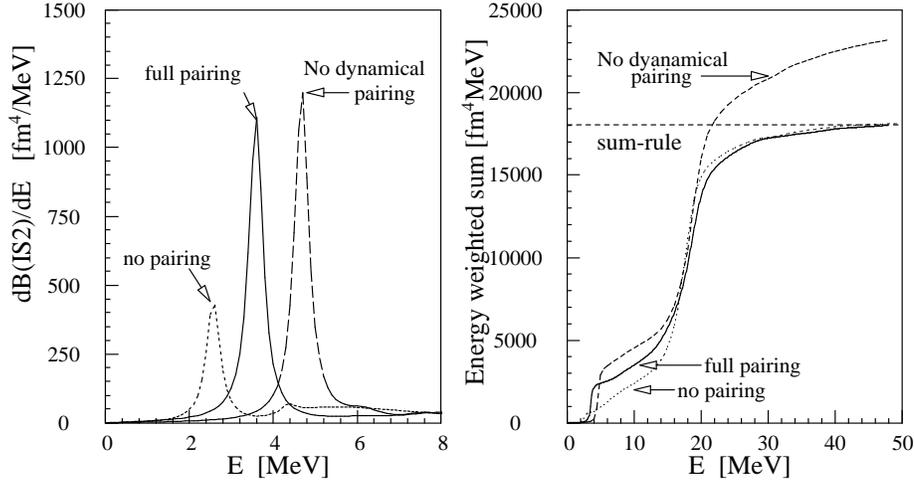}
}
\caption{Isoscalar quadrupole strength function in $^{24}$O (left) 
and its energy weighted sum (right). The calculation with the
full pairing effects (solid) is compared with the one with no pairing 
effects (dotted), and the one where 
only the static HFB potential is taken into account and 
the dynamical pairing effect is neglected (dashed). }
\label{fig:5}
\end{figure}

Effects of neutron pairing on the
quadrupole response are quite large.
The pairing correlation increases drastically the collectivity
of the low-lying
$2^+$ state, as shown in Fig.5 (solid vs. dotted curves).
The energy weighted isoscalar sum 
$S({\rm IS}2)=\int E \left({dB({\rm IS}2)\over dE}\right) dE$ 
of the $2^+$ state (and the E2 sum $S({\rm E}2)$) is calculated to be
$ 730, 1220, 1680, 2330  \ {\rm fm}^4{\rm MeV}$
(and $ 43, 47, 54, 62  \ e^2{\rm fm}^4{\rm MeV}$ ) 
for $A=18-24$, while by neglecting the pairing correlations
these quantities become 
$ 73, 132, 277, 646 \ {\rm fm}^4{\rm MeV}$ (and
$ 3, 4, 7, 14$   $e^2{\rm fm}^4{\rm MeV}$), which are
several times smaller.
Here we emphasize 
that the total pairing effect arises not only from 
the static HFB mean-field but also from the dynamical RPA
correlation induced by the pairing interaction.  
As Fig.5 (left) indicates,
the dynamical pairing correlation lowers the excitation energy
of the $2^+$ state by  $\approx 1$ MeV  (solid vs. dashed curves). 

The dynamical pairing correlation has another important role as shown in 
Fig.5 (right), where accumulated energy weighted
isoscalar sum $\int_0^E E' \left({dB({\rm IS}2)\over dE'}\right) dE'$
is plotted. Note that
the energy weighted sum rule
is satisfied quite accurately, but this
\begin{wrapfigure}{r}{6.6cm}
\centerline{
 \epsfxsize = 50mm\epsfbox{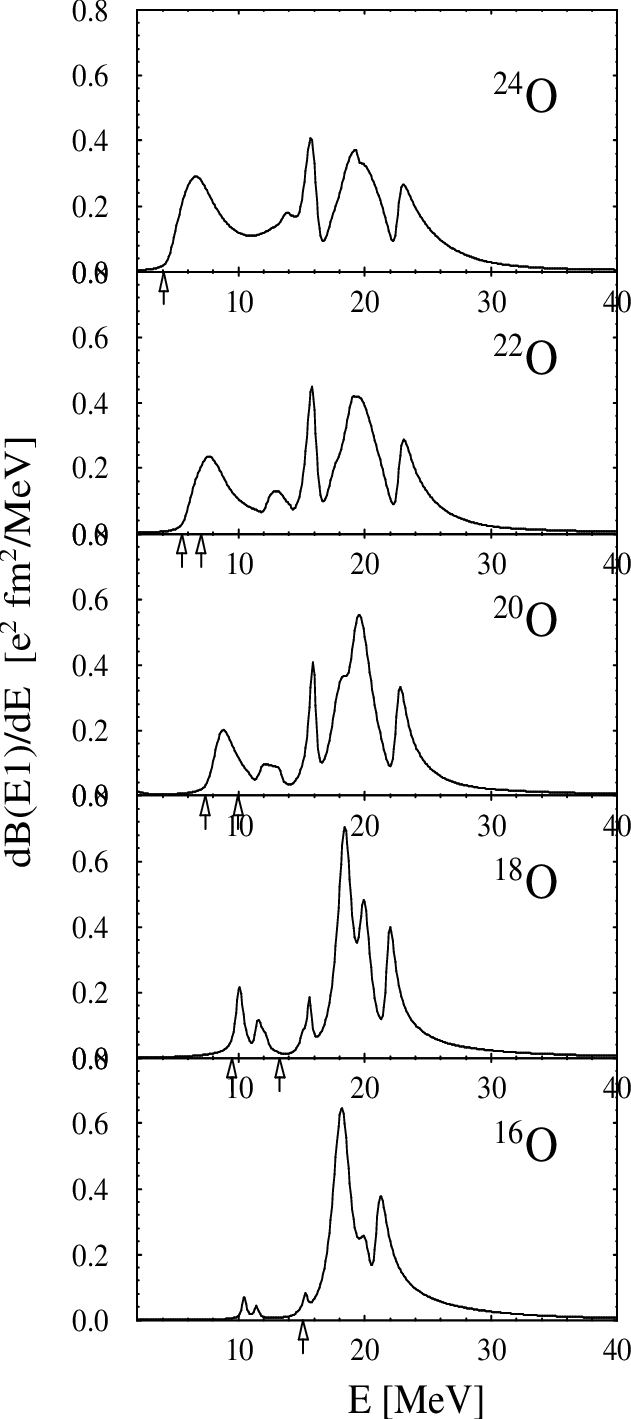}
}
\caption{Electric dipole strength function for $^{16-24}$O.
The arrows indicate the one- and two-neutron thresholds.}
\label{fig:6}

\vspace{10mm}

\centerline{
\epsfxsize=60mm\epsfbox{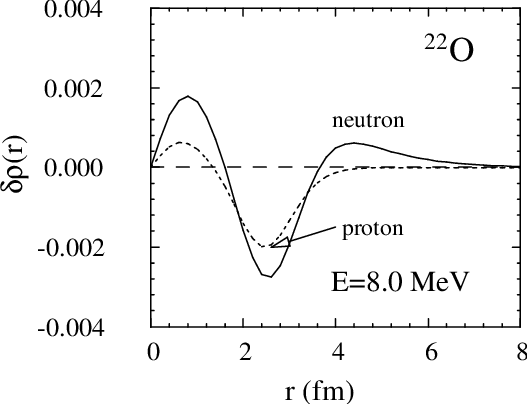}
}
\caption{The transition density for the E1 response 
of neutron (solid) and proton (dashed) at $E=8.0$ MeV in $^{22}$O,
plotted as a function of the radial coordinate $r$.}
\label{fig:7}
\end{wrapfigure}
 is achieved only by including selfconsistently
the dynamical pairing correlation on top of
the mean-field pairing effect.
If one neglects the dynamical pairing,
the selfconsistency in the linear response equation is 
broken and the sum rule is violated as seen for the dashed curve.

\subsection{Dipole excitation}

The dipole excitation is very interesting since 
the microscopic structure of the low-energy E1 mode,
called often the soft dipole mode 
or the pygmy dipole resonance, 
is currently debated\cite{HaSaZh,Hansen,Suzuki,Catara,Ring,GSI}
while a large effect of the pairing correlation 
on this mode is also pointed in the case of the halo 
nucleus $^{11}$Li.\cite{Esbensen}
We here
calculate the electric dipole strength by using
the operator $D_\mu=e{Z\over A}\sum_n rY_{1\mu}(n) -e
{N\over A}\sum_p rY_{1\mu}(p)$, 
from which the center of mass motion is explicitly removed.
The result is shown in Fig.6. As the neutron number increases
the E1 strength in the low energy region (e.g. $E < 15$ MeV) develops
significantly; the energy weighted sum below 15 MeV is
$S({\rm E}1)=7, 11, 16, 21\%$ of the TRK sum-rule value 
for $A=18-24$. 
The transition density $\delta\rho(r)$ (shown in Fig.7) indicates
that only neutrons are moving in the exterior region,
whereas in the surface region neutrons and protons move coherently
in the direction opposite to the outside neutrons. 
Thus the low-energy structure 
has at least partly the character 
of the pygmy resonance 
or the soft dipole mode.\cite{Suzuki,Hansen}
The low-energy strength may also be related to 
the so called threshold strength since the strength increases sharply
just above the one-neutron threshold $E_{th,1n}$ (see Fig.6).
In fact, it cannot 
be described well by discretizing the neutron continuum states 
with use of, e.g.
a spherical box boundary condition with a box radius $R=20$ fm
since this is a structure consisting of genuine continuum states.
Note however that the low energy strength does not arise purely from
unperturbed neutron continuum states, as we see below.

Neutron pairing effect on the E1 strength is strong as shown 
in Fig.8, where different contributions of the effects are also examined.
The total pairing effect enhances the low-energy E1 strength
near the threshold $E\sim 8$MeV. 
It is also seen that the static mean-field pairing
increases the low-energy strength  with respect to the
unpaired response (dashed vs. dotted curves). This effect may be attributed 
to partial filling, due to the pairing, of the neutron $2s{1\over 2}$ orbit
which is influential on the threshold strength.  
However, more significant is 
the enhancement caused by the dynamical pairing correlation
(solid vs. dashed curves).
This indicates that 
the enhanced low-energy strength is not
simply the threshold strength associated with the unperturbed
quasi-neutron states occupying the continuum and the weakly bound orbits,
but rather it is due to the pairing RPA correlation that mixes different
two-quasineutron states. A similar effect is seen 
in Ref.\citen{Esbensen} in their analysis of $^{11}$Li using
a two-neutron continuum shell model.

\begin{figure}[t]
\centerline{
\epsfxsize=90mm\epsfbox{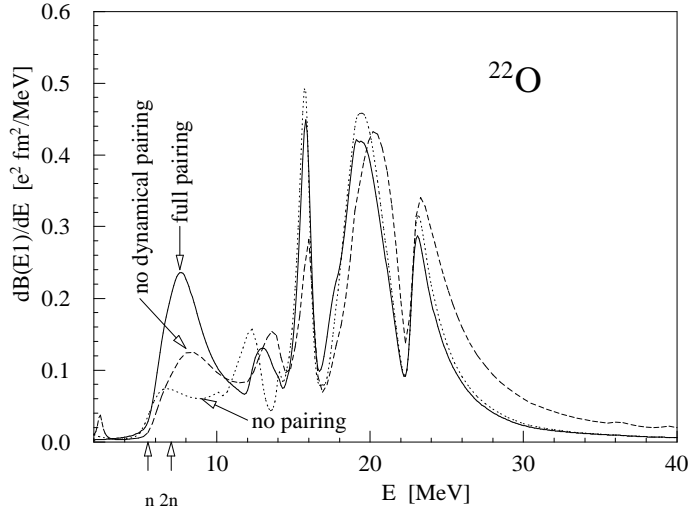}
}
\caption{Effects of the pairing correlations 
on the dipole response in $^{22}$O.
The solid curve is obtained with full pairing effects included.
For the dotted curve no pairing effect is included
whereas in the calculation shown by the dotted curve 
only the static HFB potential is taken into account and 
the dynamical pairing effect is neglected. 
The TRK sum rule is satisfied for the solid and dotted curves, but not for
the dashed curve.}
\label{fig:8}
\end{figure}

\section{Conclusions}

The new linear response theory formulated in the coordinate-space HFB enables
us to describe the collective excitations in nuclei near drip-line where
the pairing correlation and the coupling to continuum states are important.
The dynamical RPA correlations in the excited states
are taken into account both in the ph- and pp-channels 
in a way consistent with the description of the static HFB mean-field. 
The continuum states in the one- and two-particle escaping channels 
are included. Since the theory satisfies the pairing
selfconsistency both in the HFB ground state and in the linear
response equation, 
there is no spurious excitation of nucleon number and the
energy weighted sum rule is guaranteed with good accuracy.
The dynamical pairing correlation in the pp-channel 
causes strong enhancement in the strength of
the low-lying quadrupole neutron vibration
and of the low-energy dipole excitation as demonstrated with
the numerical analysis for
oxygen isotopes near the neutron drip-line.


\end{document}